# Unusual charge density wave introduced by Janus structure in monolayer vanadium dichalcogenides

Ziqiang Xu[1]†, Yan Shao[1]†, Chun Huang[1]†, Genyu Hu[1], Shihao Hu[1], Zhi-Lin Li[2], Xiaoyu Hao[1], Yanhui Hou[1], Teng Zhang[1], Jin-An Shi[2], Chen Liu[3], Jia-Ou Wang[3], Wu Zhou[2], Jiadong Zhou[1], Wei Ji[4], Jingsi Qiao[1]*, Xu Wu[1]*, Hong-Jun Gao[2], and Yeliang Wang[1]*

[1]School of Integrated Circuits and Electronics and Advanced Research Institute of Multidisciplinary Sciences & School of Physics, Beijing Institute of Technology, Beijing 100081, China
[2]Institute of Physics and University of Chinese Academy of Sciences, Chinese Academy of Sciences, Beijing, 100190, China
[3]Institute of High Energy Physics, Chinese Academy of Sciences, Beijing 100049, China
[4]Beijing Key Laboratory of Optoelectronic Functional Materials & Micro-Nano Devices, Department of Physics, Renmin University of China, Beijing 100872, China
†These authors contributed equally to this work.
*Corresponding author. Email: qiaojs@bit.edu.cn (J.Q.), xuwu@bit.edu.cn (X.W.), yeliang.wang@bit.edu.cn (Y.W.)

## Abstract

As a fundamental structural feature, the symmetry of materials determines the exotic quantum properties in transition metal dichalcogenides (TMDs) with charge density wave (CDW). Breaking the inversion symmetry, the Janus structure, an artificially constructed lattice, provides an opportunity to tune the CDW states and the related properties. However, limited by the difficulties in atomic-level fabrication and material stability, the experimental visualization of the CDW states in 2D TMDs with Janus structure is still rare. Here, using surface selenization of $VTe_2$, we fabricated monolayer Janus VTeSe. With scanning tunneling microscopy, an unusual $\sqrt{13} \times \sqrt{13}$ CDW state with threefold rotational symmetry breaking was observed and characterized. Combined with theoretical calculations, we find this CDW state can be attributed to the charge modulation in the Janus VTeSe, beyond the conventional electron-phonon coupling. Our findings provide a promising platform for studying the CDW states and artificially tuning the electronic properties toward the applications.

**INTRODUCTION**

Symmetry, a fundamental feature, plays a significant role in determining the properties of materials. As a collective excitation of atomic lattice, the charge density wave (CDW) transition modulates the symmetry and the electron density, especially for cases with unusual order. In electron-correlated two-dimensional (2D) transition metal dichalcogenides (TMDs), the CDW states always exist and are associated with their exotic quantum properties, such as Mott insulating, spin density wave (SDW), 2D magnetism, and unconventional superconductivity(*1-12*). Thus, manipulating the symmetry of the lattice and the CDW states, such as breaking the rotational symmetry, offers extraordinary opportunities to tune their many-body ground states to introduce new physics and applications(*13-17*). Breaking the lattice inversion symmetry, the Janus structure-an artificially constructed lattice, provides unprecedented possibilities to tune electronic states by engineering structural asymmetries. Recently, Janus structures have been realized on semiconducting 2D TMDs and brought distinct properties such as strong Rashba spin splitting and enhanced piezoelectric effect(*18-21*). However, limited by both their fabrication controllability and stability, the metallic 2D TMDs with Janus structure and their CDW states are rarely studied, leaving only a few theoretical predictions(*22, 23*).

As representative members in metallic TMDs, Vanadium dichalcogenides ($VX_2$, X=S, Se, and Te) always own CDW states. They are expected to host plentiful structures with exotic properties, such as the in-plane anomalous Hall effect in the $VS_2$-VS superlattice with symmetry breaking, as reported in our previous work(*24*). Associated with the debatable 2D magnetic properties, 2D $VX_2$ systems show distinct layer-dependent CDW states with reducing dimensionality (*25-29*). Specifically, the monolayer $VS_2$ and $VSe_2$ show various CDW superlattices with rotational symmetry breaking, such as $\sqrt{7} \times \sqrt{3}$ (*29, 30*), while the monolayer $VTe_2$ owns a modulated $4 \times 4$ CDW order, following the symmetry of its Bragg lattice(*31, 32*). These CDW states with different periods and symmetries imply that multiple driven forces may contribute to the formation of CDW, in addition to the conventional electron-phonon coupling. Introducing the Janus structure is expected to artificially change the dipole moment and the corresponding band structure of the 2D vanadium dichalcogenides, providing a promising platform for investigating the CDW states.

In this paper, we report the unusual CDW states with rotation symmetry breaking in artificial monolayer Janus VTeSe (J-VTeSe). By atomic-scale selenization of top-layer Te atoms in monolayer VTe$_2$, J-VTeSe was fabricated and determined by X-ray photoelectron spectroscopy (XPS) and scanning transmission electron microscopy (STEM). Then, using scanning tunneling microscopy/spectroscopy (STM/STS), a rhombic $\sqrt{13} \times \sqrt{13}$ CDW state was characterized in atomic scale. The energy- and spatial-resolved STS results reveal the distinct and complicated orbital texture near the Fermi energy of monolayer J-VTeSe. Combined with density functional theory (DFT) calculations, we suggest that the $\sqrt{13} \times \sqrt{13}$ CDW state in the J-VTeSe can be attributed to the charge modulation, beyond the conventional electron-phonon coupling. By introducing the Janus structure and modifying the CDW orders, we provide an effective way to tune the electronic properties of the metallic TMDs materials for their future applications.

**RESULTS**

**Fabrication of monolayer Janus VTeSe**

The J-VTeSe sample was fabricated by atomic-scale surface selenization of monolayer VTe$_2$, as shown by the schematic in Fig. 1A. We prepared monolayer VTe$_2$ on the bilayer graphene/SiC (0001) substrate (left panel). Then the surface selenization process was performed by depositing Se onto the monolayer VTe$_2$, which was kept at ~ 523 K. After the selenization process, the region with J-VTeSe is formed (right panel), and the ratio of this region can be increased by lengthening the selenization time, which can be attributed to Se atoms gradually substituting top-layer Te atoms of the VTe$_2$ structure.

To characterize the selenization of monolayer VTe$_2$, we performed the in-situ XPS measurements on the intrinsic and selenized VTe$_2$ samples. As shown in Fig.1B (more information see figs. S1 and S2 in the supplementary material), the doublet peaks originating from the V 2p$_{1/2}$ and 2p$_{3/2}$ core levels of the intrinsic VTe$_2$ structure (colored in blue) are located at 519.93 / 512.41 eV, respectively. After selenization for 30 mins, 60 mins, and 360 mins, the peaks shift to higher binding energy at 520.23 / 512.60, 520.27 / 512.74, and 520.40 / 512.81 eV, respectively. Combined with the XPS results of VSe$_2$(*27*), the energy shift of ~0.40 eV of the selenized VTe$_2$ confirms the selenization and indicates the change of the gradual structural transformation. In addition, the doublet peaks originating from the Te 3d and Se 3d core levels of the selenized sample both show an energy shift of ~0.10 eV compared to the intrinsic VTe$_2$ and VSe$_2$, respectively. The uniform energy shift of the selenized sample also indicates the changes in chemical states after the selenization (fig. S3), though the value is close to the energetic resolution.

**Characterizing atomic structure of monolayer vanadium dichalcogenides**

To determine the atomic structure of the selenized vanadium dichalcogenides sample in atomic scale, we conducted STEM measurements on the samples selenized for 60 mins. The large-scale bright-field and dark-field cross-section images (fig. S4) demonstrate a well-ordered sandwiched-like structure(*33*). As marked by the model, a typical 1T structure was confirmed by the zoomed-in annular dark-field (ADF) STEM results, as shown in Fig. 1C. Moreover, the element information of atoms can be further distinguished by the corresponding real-space atomic resolution electron energy loss spectroscopy (EELS) elemental map (Fig. 1D). Unambiguously, the results demonstrate the sandwiched-like structure, from top to bottom, is consist of Se, V and Te atoms with one atomic-layer thickness, as colored by red, green and blue, respectively. Therefore, we can confirm that the J-VTeSe has been fabricated by direct selenization of monolayer $VTe_2$ with atomic-level control.

Following the atomic characterization in the side-view, we performed STM measurements to visualize the top-view atomic structure. STM images of the samples with different selenization time were taken at room temperature (fig. S5). Without CDW superlattice, the atomic-resolutions STM images on different regions reveal the close-packed structure. The lattice constant difference between $VTe_2$ and J-VTeSe is very small, which are of 0.36 nm and 0.35 nm, corresponding to $VTe_2$ and J-VTeSe, respectively. To further characterize the atomic structure, low-temperature STM measurements (4.5 K) were performed on the sample selenized for 60 mins. In Fig. 2A, we can see two kinds of regions of the sample, and the height difference is ~0.06 nm, according to the height profile in Fig. 2B. The height difference between $VTe_2$ and J-VTeSe follows their atomic models. Unlike the results of room-temperature STM, the J-VTeSe demonstrates a distinct superlattice from $VTe_2$ in the low-temperature (LT) STM image (fig. S6), which could be attributed to the CDW orders(*31, 32*).

**Atomic structure and electronic states of CDW**

To investigate the CDW states, we obtained atomic-resolution STM images of monolayer $VTe_2$ and J-VTeSe, using LT-STM at 4.5 K. As shown in Fig. 2C, a typical CDW superstructure with threefold rotational symmetry can be observed in the monolayer $VTe_2$. The STM image indicates a $4 \times 4$ CDW order with the lattice constant $a_{cdw}$=1.42 nm (marked as black arrows), consistent with previous reports (*32*). Owning three-fold symmetry, two sets of spots can be observed in the corresponding fast Fourier transform (FFT) pattern (shown in Fig. 2C inset). The outer six spots represent the Bragg lattice of $VTe_2$ (black circles), and the inner six spots represent the CDW lattice (cyan circles), indicating the hexagonal $4 \times 4$ CDW superstructure.

Distinct from $VTe_2$, J-VTeSe owns a rhombic $\sqrt{13} \times \sqrt{13}$ CDW order with threefold rotation symmetry breaking, as shown in Fig. 2E. The unit cell of the superlattice is marked by the black rhombic with the lattice constant $a_{cdw}$=1.26 nm ($\sqrt{13} \times a_{VTeSe}$=1.26 nm). Moreover, the rhombic $\sqrt{13} \times \sqrt{13}$ has an angle between the two basic vectors of ~88°, which is distinct from the well-known star of David $\sqrt{13} \times \sqrt{13}$ R13.9° superlattice(*11, 12*). From the corresponding FFT pattern (as inset), we can identify six spots of the J-VTeSe Bragg lattice (black circles), which are similar to $VTe_2$. Unlike the results of $VTe_2$, it shows another four spots in the pattern of J-VTeSe, indicating the rhombic CDW superlattice (purple circles). These FFT spots further verify the rhombic $\sqrt{13} \times \sqrt{13}$ CDW orders.

Considering the CDW superlattice of $VTe_2$ and J-VTeSe, the corresponding atomic models can be labeled, as shown in Figs. 2D and 2F. Compared with the $4 \times 4$ CDW order of $VTe_2$, the $\sqrt{13} \times \sqrt{13}$ CDW superlattice in J-VTeSe could be regarded as the two basic vectors rotating outward for ~14°. The terminal of the $\sqrt{13} \times \sqrt{13}$ basic vectors can be considered as those of the $4 \times 4$ superlattice moving for 1 unit cell along the atomic lattice as shown by the rhombuses in Fig. 2F Then each $\sqrt{13} \times \sqrt{13}$ unit cell owns 15 V atoms, which is 1 atom less than the $4 \times 4$ unit cell. Although both of them own the same V atom number, it is worth noting that the $\sqrt{13} \times \sqrt{13}$ CDW of J-VTeSe is different from the "star-of-David" $\sqrt{13} \times \sqrt{13}$ CDW order in $TaSe_2$ or $NbSe_2$, the latter of which still keeps the threefold symmetry (*9-12*). As a result, it formed an unusual CDW order with two-fold symmetry instead of the three-fold symmetry $4 \times 4$ CDW order of $VTe_2$, implying other driving forces should contribute to the CDW formation.

To investigate the electronic structure of the CDW state, we performed energy- and spatial-resolved STS measurements to study the electronic local density of states (LDOS) of J-VTeSe. The typical dI/dV spectra taken on $VTe_2$ and J-VTeSe, respectively, are shown in Fig. 3A (upper panel). Both spectra indicate the CDW gap at the Fermi level, appearing as a wide dip with a finite value at its minimum. Besides the CDW gap, the big peaks contributed by the *d* orbital of V atoms show the position at ~ -100 meV for VTe$_2$ and ~ +60 meV for J-VTeSe. The ~160 meV energy shift of the peak near Fermi energy reveals that the Janus dipoles existing in J-VTeSe will introduce doping and affect the band structures, since most characteristic peaks of these two spectra are similar.

In order to gain additional insight into the CDW states at different energies, we performed dI/dV spatial mapping of J-VTeSe at different bias voltages near Fermi energy ($E_F$) as shown in Fig. 3 (B to F) (more information see figs. S7 and S8), indicating the energy-dependent orbital textures. Considering the tip induced switch of the CDW state in high bias scanning (fig. S9), we did the STS mapping at the bias from -350 mV to 350 mV with step of 10 mV. The mappings measured at different biases show distinct features, especially for those taken on the energy below and above Fermi energy. As marked by the yellow circles, we can see the contrast inversion appears at the mapping at -100 meV and 50 meV, which indicates the superlattice origin from CDW transitions.

To further quantify the energy-dependent LDOS distribution of VTe$_2$ and J-VTeSe, we cross-correlated our dI/dV maps with a reference map taken at +300 meV of the conduction band(*11*). The resulting normalized cross-correlation values of both VTe$_2$ and J-VTeSe as a function of bias voltages are color-coded in the lower panel of Fig. 3A. The conduction band (V > 0 V) of both VTe$_2$ and J-VTeSe all have a strong and positive cross-correlation (near to 1, yellow). It means that the orbital texture remains almost unchanged at the conduction band for both VTe$_2$ and J-VTeSe.

At the energy of the valence band (V < 0 V), the normalized cross-correlation curves show different features between VTe$_2$ and J-VTeSe. The dI/dV mappings of VTe$_2$ at the energy between -300 meV to -200 meV show strongly anti-correlated with the conduction band map, while it flips to positive at -150 meV, which is still far from the Fermi energy. For the J-VTeSe, the mappings are partially anti-correlated at the energy between -100 meV to 0 meV, with the conduction band map. Specifically, it flips to the positive cross-correlation at the Fermi energy. Moreover, compared with the smooth curve of VTe$_2$, the normalized cross-correlation curve of J-VTeSe is fluctuant, which indicates the complicated orbital texture. With the combination of the rotation symmetry breaking and the unique orbital texture, we can see that the mechanism of the CDW state in J-VTeSe is beyond the simple conventional Peierls description, which induces the CDW state in VTe$_2$, and should be associated by other mechanism (*34*).

**DFT calculation results of monolayer Janus VTeSe**

To investigate the CDW mechanism of J-VTeSe, DFT calculations were performed on monolayer $VTe_2$, J-VTeSe, and $VSe_2$. The lattice constant of the J-VTeSe (3.44 Å) is in between those of $VTe_2$ (3.53 Å) and $VSe_2$ (3.30 Å) (Fig. 4 A, more detail about the $d_1$ and $d_2$ see table S1), which show a similar trend with other Janus 2D TMDs (*18, 19*). Due to the inherent asymmetry of Janus structure, a 5 pC/m dipole moment from Se to Te directions was induced into the J-VTeSe. The calculated non-spin-polarized band structures of these three materials (fig. S10), matching well with previous reported experimental results(*35, 36*). The band structures and Fermi surfaces of $VTe_2$, $VSe_2$ and J-VTeSe are shown in fig. S11 to analyse the Fermi nesting vectors. We found $q$ vector for J-VTeSe is 0.25 **g** (where **g** represents the reciprocal lattice vector), which corresponds to a $4a_0$ supercell, like the results of $VTe_2$, instead of the $\sqrt{13}$ $a_0$ superstructure in the experimental results of J-VTeSe.

Furthermore, the electron-phonon coupling plays a vital role in the CDW formation of other TMDs(*36-38*). The phonon dispersion spectrum of J-VTeSe is calculated as shown in Fig. 4B. For the J-VTeSe, the lowest imaginary frequencies were located at $q$ point (0.25, 0, 0) in the non-spin-polarized configuration. These imaginary frequencies still induce the $4a_0$ supercell, instead of $\sqrt{13}$ $a_0$ in real space. With the combination of the calculation results above, we can see that conventional Fermi surface nesting and electron-phonon coupling cannot fully account for the formation mechanism of the CDW state in J-VTeSe, especially for the rotation symmetry breaking and $\sqrt{13}$ $a_0$ superstructures. In addition, we found the corresponding phonon mode can be softened and displays Kohn anomaly with increasing temperature, only if we consider the spin-polarized configuration (fig. S12). It indicates the competition between the spin polarization and the CDW states. Then we can infer that the mechanism beyond the conventional electron-phonon coupling is involved in the formation of CDW states in J-VTeSe.

Charge modulation, induced by electron-electron interaction, *etc.*, is highly correlated with the formation of CDW phase transition *(45,48)*. To investigate the charge modulation effect in the J-VTeSe, we further analyzed the band structures and density of states in $\sqrt{13} \times \sqrt{13}$ supercell with pristine and CDW configurations. For the CDW phase, we built $\sqrt{13} \times \sqrt{13}$-88° CDW superlattice and searched for two candidates named configurations A and B (fig. S13). According to the calculation results, configuration A, even with lower local symmetry, is more stable of 0.67 meV/ unit than configuration B. Moreover, the STM simulation based on configuration A is consistent with the experimental results, as shown in Fig. 4C. Then our calculation focused on configuration A and found a relatively irregular gathering together of V and Se atoms, especially compared with the well-known $\sqrt{13} \times \sqrt{13}$ star-of-David CDW of TMDs, e.g. 1T-NbSe$_2$ (*12*), which may associate with the unique electronic structures.

Furthermore, we calculated the band structures of the J-VTeSe with pristine and CDW configurations. For the electronic states at Gamma point around Fermi level, both the V $d_{z^2}$ and Se $p_z$ orbital contribute to the bands around the Fermi level of J-VTeSe, combined with the feature in VSe$_2$ and VTe$_2$ shown in Fig. 4A. Therefore, we calculated the projected density of states (PDOS) of these two orbitals, as shown in Figs. 4D and 4E. Most notably, a pronounced peak is observed on the Fermi level in the PDOS of the pristine phase, corresponding to the flat band contributed by V $d_{z^2}$ and Se $p_z$ orbitals (see fig. S14 for more detail). Meanwhile, for the CDW phase, the DOS around the Fermi level is dramatically decreased. Moreover, these two orbitals become delocalized and disperse into several bands away from the Fermi level. This can be attributed to the charge modulation with the induced dipole and V $d_{z^2}$/Se $p_z$ hybridization, which disrupts the usual band dispersion. Then the system can be transformed from a high-symmetry initial state to a low-symmetry new ground state, which may be the mechanism of the symmetry-breaking CDW state in J-VTeSe.

**SUMMARY**

In summary, we have introduced the Janus structure into 2D vanadium dichalcogenides and artificially fabricated J-VTeSe by atomic-scale direct selenization. By various atomic-level characterization methods, we confirmed the atomic structure of J-VTeSe and its newly emerged CDW order with threefold rotation symmetry breaking. The spectroscopic imaging experimentally reveals the unusual and complicated orbital texture near Fermi energy. Combined with theoretical calculations, we found that the charge modulation is involved in the CDW formation of the 2D vanadium dichalcogenides, and acts as the assistance of conventional Peierls description. By introducing the Janus structure, we realize tuning the rotational symmetry of CDW states in electron-correlated materials. Our findings provide an effective method to artificially construct electron-correlated systems and tune their properties to the applications in multifunctional electronics.

**MATERIALS AND METHODS**

**Sample preparation**

Monolayer $VTe_2$ was epitaxial grown on bilayer graphene/SiC (0001) substrate in an ultrahigh vacuum (UHV) chamber, with a base pressure of $2\times10^{-10}$ mbar, equipped with standard MBE capabilities. The bilayer graphene/SiC (0001) substrate was prepared by annealing the doped SiC crystalline substrate (TankeBlue) at 1500 K for 180 minutes after degassing at 900 K for 5 hours. Tellurium clusters (Sigma, 99.999%) were evaporated from a Knudsen cell, vanadium atoms (ESPI Metals, 99.999%) were evaporated from an electron-beam evaporator, and they were deposited onto the bilayer graphene/SiC (0001) substrate, kept at 550 K. The growth parameters were under Te-rich conditions, with aiming to guarantee there are enough Te atoms involving in reaction with vanadium. The excessive tellurium atoms will be desorbed from the substrate since the substrate temperature at 550 K was higher than the evaporation temperature of Tellurium atoms (520 K). For the fabrication of J-VTeSe, selenium clusters (Sigma, 99.999%) were evaporated from a Knudsen cell onto the monolayer $VTe_2$, which was kept at 523 K.

**XPS measurements**

The in-situ X-ray photoelectron spectroscopy measurements were performed at the Beijing Synchrotron Radiation Facility (BSRF). Synchrotron radiation light was monochromated by four high-resolution gratings and controlled by a hemispherical energy analyzer has photon energy in the range from 10 to 1,100 eV.

**STEM measurements**

Before the STEM measurements, we firstly deposited 20 nm $C_{60}$ and 80 nm Sb films on as-grown monolayer J-VTeSe sample in UHV chamber, aiming to protect the sample from oxidation and damage. Then the samples were sliced along SiC (11-20) face by a focused ion beam (FIB) and were further thinned to around 40 nm thickness using low-energy ion milling. And the cross-section high-angle annular dark-field scanning transmission electron microscopy (HADDF-STEM) images were obtained with an aberration-corrected STEM operated at 60 kV. Experimental conditions: acceleration voltage: 60 kV; beam current: 20 pA; convergence angle: 32 mrad; HAADF collection angle: 75~210 mrad; BF collection angle: 0~10 mrad.

**Scanning tunneling microscopy measurements**

The samples were transfer into the scanning tunneling microscopy using a UHV suitcase, with the base pressure of $2\times10^{-10}$ mbar. The STM experiments of the samples were performed at both 300 K and 4.5 K. All the STM images were measured in constant current mode. All the measurements were performed with the electrochemically etched tungsten tips. The bias voltage was applied to the sample. The tunneling differential conductance (dI/dV) was measured using lock-in detection of the tunnel current by adding a 10 mV modulated bias voltage at 973 Hz of the sample bias voltage.

**First-principles calculations**

Density functional theory calculations were performed using the generalized gradient approximation for the exchange-correlation potential, the projector augmented wave method[41], and a plane-wave basis set as implemented in the Vienna ab initio simulation package (VASP)[42]. Van der Waals interactions were considered at DFT-DF level with the optB86b-vdW function[43]. The finite displacement method implemented in VASP and PHONOPY[44] was used to calculate phonon-dispersion for the monolayer J-VTeSe. The kinetic energy cutoff for the plane-wave basis set was set to 700 eV for calculating geometric and electronic properties. For the primitive unit cell of monolayer $VTe_2$, J-VTeSe and $VSe_2$, a $k$-mesh of 21×21×1 was adopted to sample the first Brillouin zone during geometry optimization and the electronic calculations. The k-mesh density of 0.004 Å$^{-1}$ in reciprocal space was utilized in the 2D fermi surface calculations. The 2D fermi surfaces were plotted at the energy range of [$E_F$ = -30 meV, $E_F$ = +30 meV]. In addition, on-site Coulomb interactions were considered on the $d$ orbitals of V atoms with effective value $U$ = 2.2 eV. All atoms in the supercell were allowed to relax until the residual force per atom was less than 5×10$^{-3}$ eV Å$^{-1}$ and the energy convergence criteria was 1×10$^{-5}$ eV.

**SUPPLEMENTARY MATERIALS**

Supplementary material is available in the online version of the paper.

FIGURES

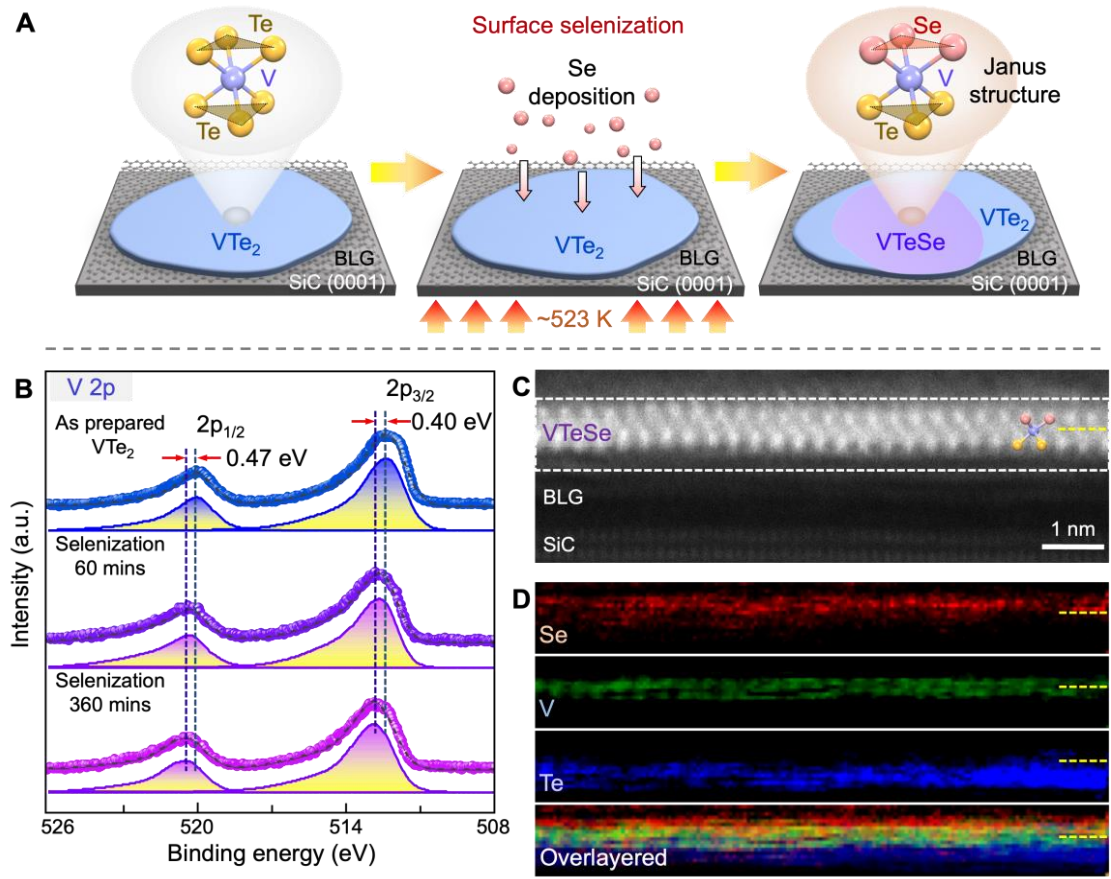

**Fig.1. Fabrication of the monolayer Janus VTeSe.** (**A**) The schematic of the synthesis process. The monolayer J-VTeSe is fabricated by surface selenization of monolayer $VTe_2$. (**B**) XPS results of V 2p spectra of the sample during the selenization process. The V 2p peak positions (of as prepared $VTe_2$ at 519.93 and 512.41 eV, of the sample with 60mins selenization at 520.27 and 512.74 eV, and of the sample with 360 mins selenization at 520.40 and 512.81 eV) show apparent energy differences before and after selenization. (**C**) ADF STEM image of the sample cross-section, showing the sandwiched-like 1T structure of the J-VTeSe monolayer with top Se (red), bottom Te (yellow), and middle V(blue) atoms. (**D**) The corresponding EELS maps of the dashed frame in (C). The layers of Se, Te, and V are colored in red, blue, and green, respectively. The yellow dashed lines mark the position of the V layer in each EELS map.

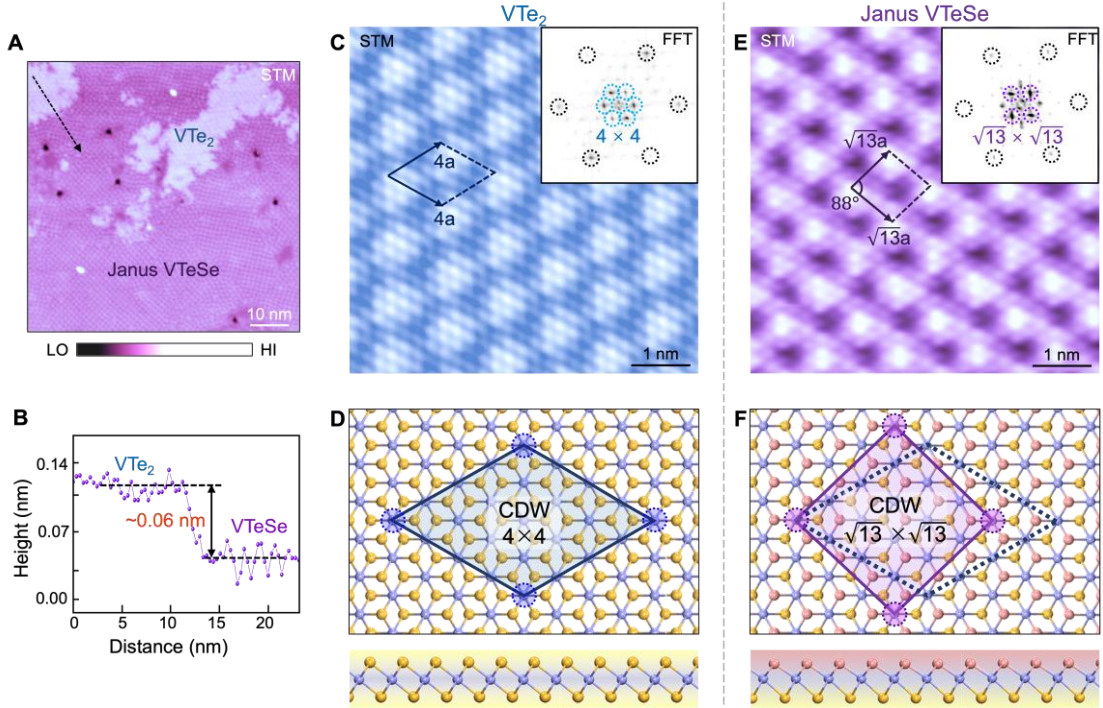

**Fig.2. Atomic structure of monolayer V dichalcogenides.** (**A**) Typical STM image taken at ~4.5 K after sample with 60mins selenization. (**B**) The corresponding height profile alongs the black arrow in (A), indicating the height difference between the $VTe_2$ and J-VTeSe. (**C** and **E**) The atomic-resolution STM images of the monolayer $VTe_2$ (C) and J-VTeSe (E), respectively. The unit cells of the CDW superstructure of $VTe_2$ and J-VTeSe are marked by black rhomboids, respectively. Inset: The corresponding FFT patterns, verifying the 4×4 CDW superlattice in $VTe_2$ (C) and the √13×√13 CDW superlattice in J-VTeSe (E), respectively. (**D** and **F**) Corresponding atomic models of $VTe_2$ (D) and J-VTeSe (F) structures in top views (upper panels) and side views (lower panels). The Se, Te, and V atoms are represented as red, yellow, and blue balls, respectively. The unit cells of the two CDW superlattices are marked and shaded for better comparison. STM parameters: Sample bias U = -1.0 V, Setpoint I = 10 pA (A); U = -0.3 V, I = 100 pA (C and E).

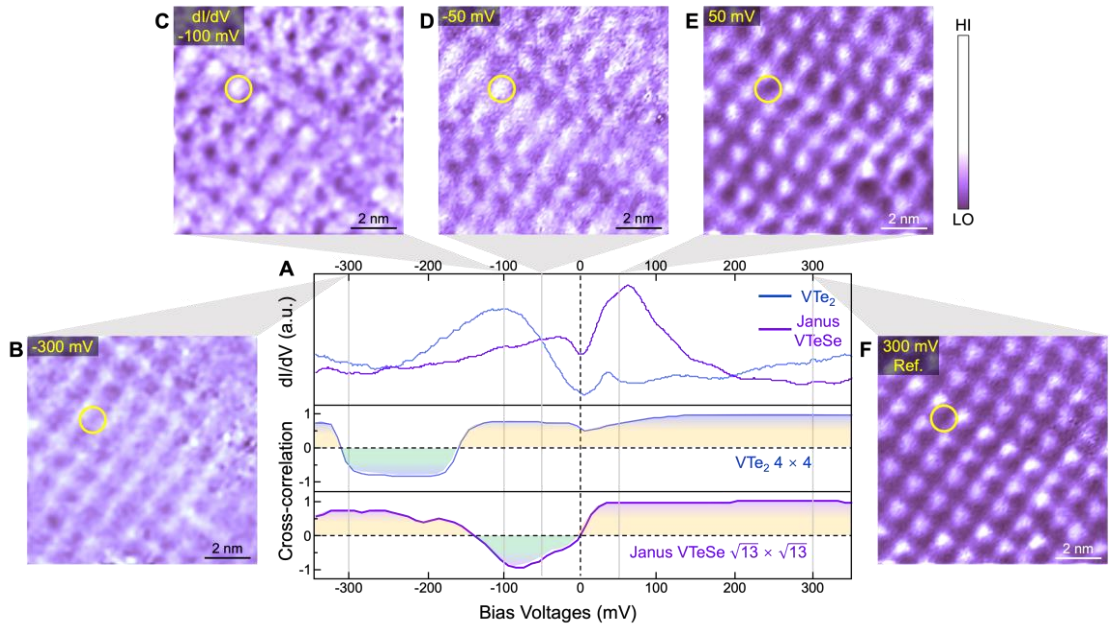

**Fig.3. Energy- and spatial-resolved STS results of unusual CDW state in monolayer Janus VTeSe.** (**A**) Upper: The STM dI/dV spectra of the monolayer $VTe_2$ and J-VTeSe, respectively. They show the CDW gap and the electronic states near the Fermi level. Middle and lower: The cross-correlation curves of dI/dV maps of both $VTe_2$ (middle) and J-VTeSe (lower) at different energies, with the reference map taken at 300 mV (shown in fig. S8 and Fig. 3F), respectively. (**B** to **F**) Constant-height dI/dV conductance maps taken at selected bias voltages. The same area is marked with yellow circles on each map. STM dI/dV parameters: modulation voltage $V_{r.m.s.}$ = 10 mV, frequency f = 973 Hz, stabilization position, U = -0.35 V, I = 50 pA.

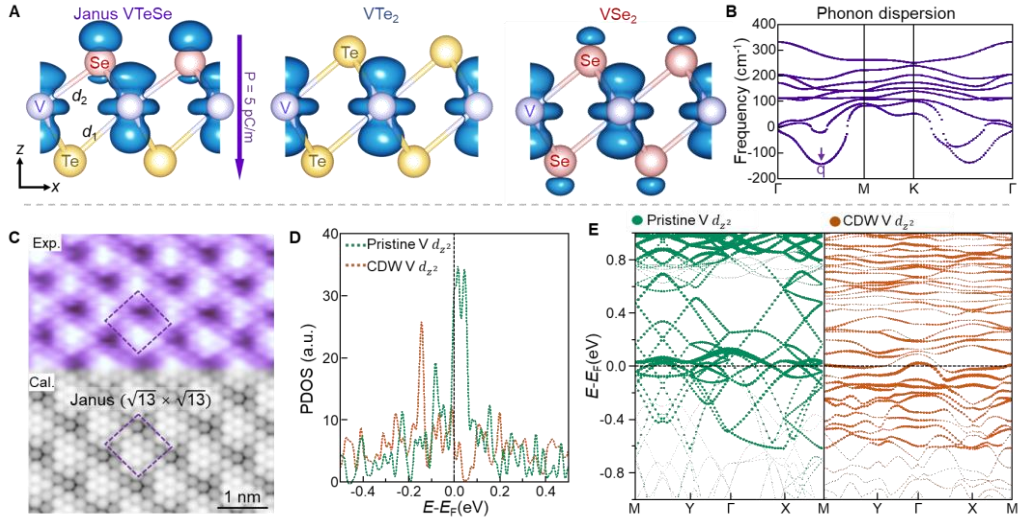

**Fig.4. The theoretical electronic structure of the monolayer VTe$_2$ and Janus VTeSe.** (**A**) The decomposed charge density for the marked S1 state (marked in the band structures in fig. S10). The models are illustrated in the *xz* planes using an isosurface of $8\times10^{-3}$ e/bohr$^3$ for J-VTeSe (left), VTe$_2$ (middle), and VSe$_2$ (right), respectively. (**B**) Phonon dispersion of monolayer J-VTeSe. The lowest imaginary frequency is located at q point (0.25, 0, 0) (marked with a purple arrow). (**C**) The experimental (upper) and simulated STM image (lower) of monolayer J-VTeSe. Purple dashed rhomboids mark the CDW superstructures. (**D**) Projected band structures of the V $d_{z^2}$ orbitals for the pristine structure (green) and CDW structure (orange) using $\sqrt{13}\times\sqrt{13}$ supercells. (**E**) PDOS of the V $d_{z^2}$ orbitals in the pristine structure (green) and CDW structure (orange).